\documentclass[runningheads,a4paper]{llncs}

\usepackage{amssymb}
\usepackage{graphicx}
\usepackage[colorlinks=true]{hyperref}
\usepackage{url}
\urldef{\mailsa}\path|yongwanl@usc.edu, shri@ee.usc.edu, knayak@usc.edu|
\begin{document}

\mainmatter  % start of an individual contribution

% first the title is needed
\title{Attention-gated convolutional neural networks for off-resonance correction \\of spiral real-time MRI\thanks{This work has been presented at the ISMRM 2020} }

% a short form should be given in case it is too long for the running head
\titlerunning{AG-CNN for off-resonance correction of spiral RT-MRI}

% the name(s) of the author(s) follow(s) next
%
% NB: Chinese authors should write their first names(s) in front of
% their surnames. This ensures that the names appear correctly in
% the running heads and the author index.
%
\author{Yongwan Lim%
% \thanks{Please note that the LNCS Editorial assumes that all authors have used
% the western naming convention, with given names preceding surnames. This determines
% the structure of the names in the running heads and the author index.}%
\and Shrikanth S. Narayanan \and Krishna S. Nayak }
\authorrunning{Y. Lim et al}
% (feature abused for this document to repeat the title also on left hand pages)

% the affiliations are given next; don't give your e-mail address
% unless you accept that it will be published
\institute{Ming Hsieh Department of Electrical and Computer Engineering\\
  University of Southern California, Los Angeles, CA, USA \\
\mailsa}
% \mailsb\\
% \mailsc\\
% \url{http://www.springer.com/lncs}}
%
% NB: a more complex sample for affiliations and the mapping to the
% corresponding authors can be found in the file "llncs.dem"
% (search for the string "\mainmatter" where a contribution starts).
% "llncs.dem" accompanies the document class "llncs.cls".
%

\toctitle{Lecture Notes in Computer Science}
\tocauthor{Authors' Instructions}
\maketitle

\begin{abstract}
Spiral acquisitions are preferred in real-time MRI because of their efficiency, which has made it possible to capture vocal tract dynamics during natural speech. A fundamental limitation of spirals is blurring and signal loss due to off-resonance, which degrades image quality at air-tissue boundaries. Here, we present a new CNN-based off-resonance correction method that incorporates an attention-gate mechanism. This leverages spatial and channel relationships of filtered outputs and improves the expressiveness of the networks. We demonstrate improved performance with the attention-gate, on 1.5 Tesla spiral speech RT-MRI, compared to existing off-resonance correction methods. 
% \emph{abstract} environment.
% \keywords{We would like to encourage you to list your keywords within
% the abstract section}
\end{abstract}

\section{Introduction}

% You are strongly encouraged to use \LaTeXe{} for the
% preparation of your camera-ready manuscript together with the
% corresponding Springer class file \verb+llncs.cls+. Only if you use
% \LaTeXe{} can hyperlinks be generated in the online version
% of your manuscript.

% The \LaTeX{} source of this instruction file for \LaTeX{} users may be
% used as a template. This is
% located in the ``authors'' subdirectory in
% \url{ftp://ftp.springer.de/pub/tex/latex/llncs/latex2e/instruct/} and
% entitled \texttt{typeinst.tex}. There is a separate package for Word 
% users. Kindly send the final and checked source
% and PDF files of your paper to the Contact Volume Editor. This is
% usually one of the organizers of the conference. You should make sure
% that the \LaTeX{} and the PDF files are identical and correct and that
% only one version of your paper is sent. It is not possible to update
% files at a later stage. Please note that we do not need the printed
% paper.

% We would like to draw your attention to the fact that it is not possible
% to modify a paper in any way, once it has been published. This applies
% to both the printed book and the online version of the publication.
% Every detail, including the order of the names of the authors, should
% be checked before the paper is sent to the Volume Editors.

Blurring and signal loss due to off-resonance are the primary limitations of spiral MRI \cite{1,2,3}. In the context of speech real-time MRI (RT-MRI), off-resonance degrades image quality most significantly at air-tissue boundaries \cite{4,5,6}, which are the exact locations of interest. Blurring and signal loss is the result of a complex-valued spatially varying convolution. In order to resolve the artifact, conventional methods \cite{7,8,9,10,11,12} reconstruct basis images at demodulation frequencies and apply spatially-varying masks to the basis images to form a desired sharp image. \\ \\
Recently, convolutional neural network (CNN) approaches have shown promise in solving this spiral deblurring task  \cite{13,14}. The conventional methods require field maps \cite{7,8} or focus metrics \cite{9,11,12} to estimate the spatially-varying mask. One of the advantages of CNN is that once trained, ReLU nonlinearity provides the mask to convolution filters, enabling spatially-varying convolution \cite{15}. Since ReLU masks out the activation in an element-wise manner, it cannot exploit local spatial or channel (filter) dependency, unlike the conventional methods. \\ \\
In this work, we present a CNN-based deblurring method that adapts the attention-gate (AG) mechanism (AG-CNN) to exploit spatial and channel relationships of filtered outputs to improve the expressiveness of the networks \cite{16,17,18}. We demonstrate improved deblurring performance for 1.5 Tesla spiral speech RT-MRI, compared to a recent CNN study \cite{14}, and several conventional methods.

\begin{figure}[t]
\centering
\includegraphics[height=4.6cm]{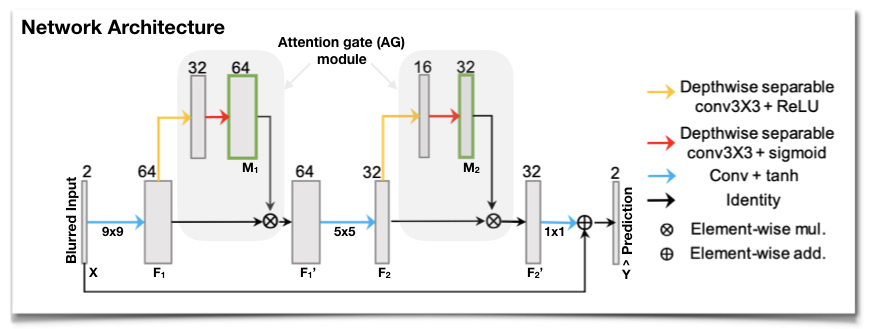}
\caption{\textbf{Network architecture}. The attention gate (AG) module is integrated with a previous CNN architecture \cite{14}. The AG consists of two cascaded depth-wise separable convolutions (each consisting of a channel-wise convolution followed by a 1$\times$1 convolution) \cite{19}, one with ReLU and one with sigmoid activations to generate attention maps ($\mathbf{M}$) in the range from 0 to 1. The attention map is then multiplied back by the convolution output ($\mathbf{F}$) (i.e.,$\mathbf{F’} = \mathbf{M(F)} \otimes \mathbf{F}$) element-by-element to emphasize important elements in space and across channels. }
\label{fig:fig1}
\end{figure}

\section{Methods}

\subsection{Network Architecture}
We use a simple 3-layer residual CNN architecture \cite{14} and incorporate a proposed AG module at each convolution layer, as illustrated in Fig.~\ref{fig:fig1}. The AG takes the output feature maps ($\mathbf{F}$) from a convolution unit as an input and performs two cascaded depth-wise separable convolutions to generate attention maps ($\mathbf{M}$) in the range from 0 to 1. Depth-wise separable convolution \cite{19} is used to improve the AG module in both performance and overhead. The attention maps $\mathbf{M}$ learn to identify salient image regions and channels adaptively for given feature maps $\mathbf{F}$, and they preserve only the activation relevant to the deblurring task in the following convolution layers. The AG multiplies the attention map by the convolution output (i.e., $\mathbf{F’} = \mathbf{M(F)} \otimes \mathbf{F}$) to emphasize important elements in space and across channels. 

\subsection{Training Data}
2D RT-MRI data from 33 subjects were acquired at our institution on a 1.5 Tesla scanner (Signa Excite, GE Healthcare, Waukesha, WI) using a vocal-tract imaging protocol \cite{20}. It uses a short readout (2.52 ms) spiral spoiled-gradient-echo sequence. Ground truth images were obtained after off-resonance correction \cite{6}. We augmented field maps estimated in the correction step by scaling $\mathbf{f'} = \alpha \mathbf{f} + \beta$ with $\alpha$ ranging from 0 to 3.15, and $\beta$ ranging from -200 to 200 Hz. Distorted images were then synthesized by using the discrete object approximation and simulating off-resonance using the field map $\mathbf{f'}$ and spiral trajectories with readout lengths of 2.520, 4.016, 5.320, and 7.936 ms. We split data into 23, 5, and 5 subjects for training, validation, and testing. 

\subsection{Network Training}
Our model was trained in a combination of L1 loss ($\mathcal{L}_1$) and gradient difference loss ($\mathcal{L}_{gdl}$) \cite{21} between the prediction and ground truth as $\mathcal{L}=\mathcal{L}_1+\mathcal{L}_{gdl}$. In addition to $\mathcal{L}_1$, $\mathcal{L}_{gdl}$ is known to provide a sharp image prediction. We used Adam optimizer \cite{22} with a learning rate of $10^{-3}$, a mini-batch size of 64, and 200 epochs. We implemented our network with Keras using Tensorflow backend.

\subsection{Experiments}
We investigate the effectiveness of the AG module by varying depth-wise separable convolution filter sizes, $f_1$ and $f_2$ of the first and second AG modules. Two cascaded convolutions in an AG module uses the same filter size of either $f_1$ or $f_2$. For comparison, we also deblur images with various existing methods: the previous CNN architecture \cite{14}, multi-frequency interpolation (MFI) \cite{7}, and iterative reconstruction (IR) \cite{23}. Note that field maps are necessary for deblurring in the latter two methods, so we assume ground truth field maps are known for those two, although those would not be available in practice. For all those methods, dynamic images were deblurred frame-by-frame. We report quantitative quality comparison using peak signal-to-noise ratio (PSNR), structural similarity (SSIM), and high-frequency error norm (HFEN). 

\begin{figure}
\centering
\includegraphics[height=6.5cm]{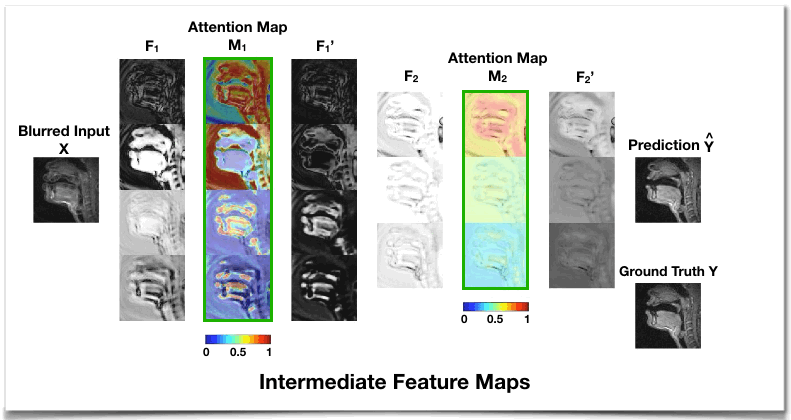}
\caption{\textbf{Intermediate feature maps. We observe that the two AG modules build hierarchical attention}. $\mathbf{M}_1$ from the first AG module tends to focus on low-level structures such as air, tissue, or air-tissue boundaries with a different focus across channels while $\mathbf{M}_2$ focuses on a high-level channel dependency. We only visualize 4 and 3 channels out of 64 and 32 channels for the first and second AG modules, respectively, due to space and file size constraints.}
\label{fig:fig2}
\end{figure}

\section{Results and Discussion}
\begin{figure}[t]
\centering
\includegraphics[height=5.8cm]{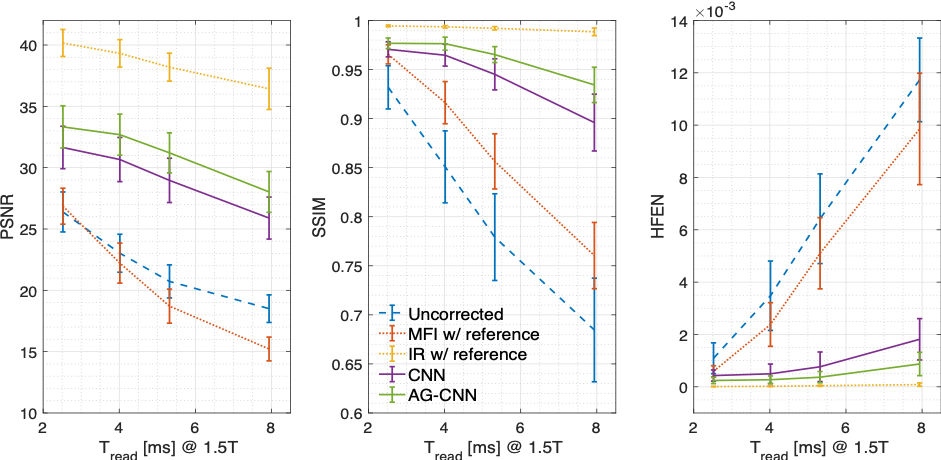}
\caption{\textbf{Quantitative comparison of deblurring performance on multiple spiral trajectories}. Four trajectories are considered with varying readout lengths of 2.52, 4.016, 5.320, and 7.936 ms. Overall, AG-CNN outperforms the previous CNN \cite{14} as well as MFI \cite{7} for PSNR, SSIM, HFEN. It should be noted that we assume a reference field map is known for both MFI and IR \cite{23} methods although it would not be available in practice. IR should be considered as an upper bound of the maximum deblurring performance achievable for a given a field map.}
\label{fig:fig3}
\end{figure}

\begin{figure}[t]
\centering
\includegraphics[height=7.3cm]{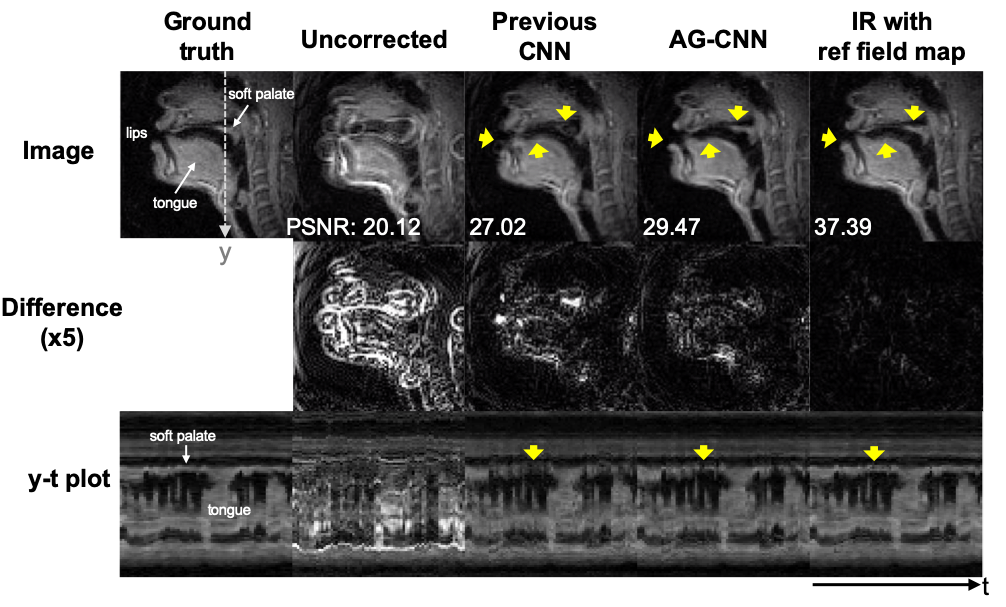}
\caption{\textbf{Qualitative comparison of deblurred images}. From top to bottom: images after deblurring with various methods, difference images with respect to ground truth, and an intensity vs time plot. The proposed AG-CNN successfully resolves the blurring artifact especially at the lips and soft palate, which is difficult to resolve with the previous CNN \cite{14}. The AG-CNN is also visually comparable to IR method \cite{23}, which uses the ground truth field map and is computationally expensive (e.g., comp. time: $\sim$ 1.6 s/frame) compared to the AG-CNN ($\sim$ 0.15 s/frame on a single CPU in inference).}
\label{fig:fig4}
\end{figure}

Fig.~\ref{fig:fig2} shows the intermediate feature maps. We observe that the attention map $\mathbf{M}_1$ from the first AG module tends to focus on low-level structures such as tissue, air, or air-tissue boundaries with a different focus across channels, while $\mathbf{M}_2$ from the second AG focuses on a high-level channel dependency.\\\\
Table \ref{tab:1} shows that adding an AG module on top of CNN layers improves deblurring performance with a slight overhead and less sensitivity to the kernel size. An extensive comparison with existing attention approaches \cite{16,17} applicable to this task remains as future work. \\\\
Fig.~\ref{fig:fig3} shows that AG-CNN outperforms the previous CNN and MFI using a reference field map in multiple readout duration lengths. Fig.~\ref{fig:fig4} contains representative image frames. Blurring of the lips and soft palate are not perfectly resolved with the previous CNN method. AG-CNN provides substantially improved depiction of these and other air-tissue boundaries. 

\begin{table}
\centering
\caption{Deblurring performance is improved by adding the proposed AG module on top of CNN layer. We obtain performance gains of $>$ 1 dB PSNR, $>$ 0.014 SSIM, and $>$ 0.029 HFEN on test dataset (5 subjects, $>$ 8K frames) with less sensitivity to the size of depth-wise separable convolution kernel in the AG module. Those evaluation metrics were averaged across all the test image frames. The number of parameters is slightly increased due to depth-wise separable convolutions in the AG module. We chose $f_1$ = $f_2$ = 3 for the rest of this study.}

\begin{tabular*}{0.8\textwidth}{@{\extracolsep{\fill} } c | c | c | c | c | c }
% \begin{tabular}{ p{2.3cm}|p{1.5cm}|p{1.5cm}|p{1.5cm}|p{1.5cm}|p{1.7cm} }
%  \multicolumn{4}{|c|}{Country List} \\
 \hline
 Architecture & ($f_1$, $f_2$) & Params &  PSNR  & SSIM & HFEN(x100)\\
 \hline \hline
 CNN(9-5-1)\cite{14}& - &  61.7K & 29.29 & 0.944 & 0.088\\
 +AG& (5,5) &  70.7K & 30.63 & 0.959 & $\mathbf{0.053}$\\
 +AG& (5,3) &  70.0K & 30.62 & $\mathbf{0.959}$ & 0.057\\
 +AG& (5,1) &  69.6K & 30.61 & 0.959 & 0.057\\
 +AG& (3,3) &  68.4K & $\mathbf{30.69}$ & 0.958 & 0.055\\
 +AG& (3,1) &  68.1K & 30.58 & 0.958 & 0.058\\
 \hline \hline
 (Blurred) Input & - & - & 22.16 & 0.812 & 0.568 \\
  \hline 

\end{tabular*}
 \label{tab:1}
\end{table}

\section{Conclusion}
We demonstrate AG-CNN deblurring for 1.5T spiral speech RT-MRI. Adding an AG module on top of CNN layer improves deblurring performance by $>$ 1dB PSNR, $>$ 0.014 SSIM, and $>$ 0.029 HFEN compared to the previous CNN architecture and provides results visually comparable to reference IR method with $\sim$ 10 times faster computation, and without the need for a field map.

\section{Acknowledgement}
This work was supported by NIH Grant R01DC007124 and NSF Grant 1514544.
\clearpage 
\end{document}